\pdfoutput=1
\documentclass[twoside,twocolumn,9pt]{article}
\usepackage{extsizes}
\usepackage[super,sort&compress,comma]{natbib} 
\usepackage[version=3]{mhchem}
\usepackage[left=1.5cm, right=1.5cm, top=1.785cm, bottom=2.0cm]{geometry}
\usepackage{balance}
\usepackage{mathptmx}
\usepackage{sectsty}
\usepackage{graphicx} 
\usepackage{lastpage}
\usepackage[format=plain,justification=justified,singlelinecheck=false,font={stretch=1.125,small,sf},labelfont=bf,labelsep=space]{caption}
\usepackage{float}
\usepackage{fancyhdr}
\usepackage{fnpos}
\usepackage{epstopdf}
\usepackage[english]{babel}
\addto{\captionsenglish}{%
  
}
\usepackage{array}
\usepackage{droidsans}
\usepackage{charter}
\usepackage[T1]{fontenc}
\usepackage[usenames,dvipsnames]{xcolor}
\usepackage{setspace}
\usepackage[compact]{titlesec}
\usepackage{hyperref}
\usepackage{physics}
\usepackage{amsmath}

\usepackage{epstopdf}

\definecolor{cream}{RGB}{222,217,201}

\begin{document}

\pagestyle{fancy}
\thispagestyle{plain}
\fancypagestyle{plain}{
\renewcommand{\headrulewidth}{0pt}
}

\makeFNbottom
\makeatletter
\renewcommand\LARGE{\@setfontsize\LARGE{15pt}{17}}
\renewcommand\Large{\@setfontsize\Large{12pt}{14}}
\renewcommand\large{\@setfontsize\large{10pt}{12}}
\renewcommand\footnotesize{\@setfontsize\footnotesize{7pt}{10}}
\makeatother

\renewcommand{\thefootnote}{\fnsymbol{footnote}}
\renewcommand\footnoterule{\vspace*{1pt}%
\color{cream}\hrule width 3.5in height 0.4pt \color{black}\vspace*{5pt}} 
\setcounter{secnumdepth}{5}

\makeatletter 
\renewcommand\@biblabel[1]{#1}            
\renewcommand\@makefntext[1]%
{\noindent\makebox[0pt][r]{\@thefnmark\,}#1}
\makeatother 
\renewcommand{\figurename}{\small{Fig.}~}
\sectionfont{\sffamily\Large}
\subsectionfont{\normalsize}
\subsubsectionfont{\bf}
\setstretch{1.125} 
\setlength{\skip\footins}{0.8cm}
\setlength{\footnotesep}{0.25cm}
\setlength{\jot}{10pt}
\titlespacing*{\section}{0pt}{4pt}{4pt}
\titlespacing*{\subsection}{0pt}{15pt}{1pt}

\fancyfoot{}
\fancyfoot[LO,RE]{\vspace{-7.1pt}\includegraphics[height=9pt]{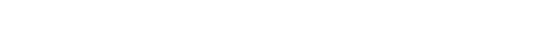}}
\fancyfoot[CO]{\vspace{-7.1pt}\hspace{11.9cm}\includegraphics{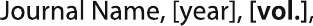}}
\fancyfoot[CE]{\vspace{-7.2pt}\hspace{-13.2cm}\includegraphics{head_foot/RF}}
\fancyfoot[RO]{\footnotesize{\sffamily{1--\pageref{LastPage} ~\textbar  \hspace{2pt}\thepage}}}
\fancyfoot[LE]{\footnotesize{\sffamily{\thepage~\textbar\hspace{4.65cm} 1--\pageref{LastPage}}}}
\fancyhead{}
\renewcommand{\headrulewidth}{0pt} 
\renewcommand{\footrulewidth}{0pt}
\setlength{\arrayrulewidth}{1pt}
\setlength{\columnsep}{6.5mm}
\setlength\bibsep{1pt}

\makeatletter 
\newlength{\figrulesep} 
\setlength{\figrulesep}{0.5\textfloatsep} 

\newcommand{\topfigrule}{\vspace*{-1pt}%
\noindent{\color{cream}\rule[-\figrulesep]{\columnwidth}{1.5pt}} }

\newcommand{\botfigrule}{\vspace*{-2pt}%
\noindent{\color{cream}\rule[\figrulesep]{\columnwidth}{1.5pt}} }

\newcommand{\dblfigrule}{\vspace*{-1pt}%
\noindent{\color{cream}\rule[-\figrulesep]{\textwidth}{1.5pt}} }

\makeatother

\twocolumn[
  \begin{@twocolumnfalse}

\vspace{1em}
\sffamily
\centering \noindent\LARGE{\textbf{Spin-momentum entanglement in a Bose-Einstein condensate}} \\
\vspace{0.3cm} \vspace{0.3cm} 

\noindent\large{Sumit Suresh Kale,$^{\ast}$ Yijue Ding,\textit{$^{\ast}$} Yong P. Chen,\textit{$^{a \dag}$} Bretislav Friedrich,\textit{$^{\ddag}$} and Sabre Kais,\textit{$^{a \ast \dag}$}}  
\vspace{0.6cm}
\begin{abstract}
    Entanglement is at the core of quantum information processing and may prove essential for quantum speed-up. Inspired by both theoretical and experimental studies of spin-momentum coupling in systems of ultra-cold atoms, we investigate the entanglement between the spin and momentum degrees of freedom of an optically trapped BEC of $^{87}$Rb atoms. We consider entanglement that arises due to the coupling of these degrees of freedom induced by Raman and radio-frequency fields and examine its dependence on the coupling parameters by evaluating von Neumann entropy as well as concurrence as measures of the entanglement attained. Our calculations reveal that under proper experimental conditions significant spin-momentum entanglement can be obtained, with von Neumann entropy of 80\% of the maximum attainable value. Our analysis sheds some light on the prospects of using BECs for quantum information applications.
\end{abstract}

 \end{@twocolumnfalse} \vspace{0.6cm}
  ]


\renewcommand*\rmdefault{bch}\normalfont\upshape
\rmfamily
\section*{}
\vspace{-1cm}


\footnotetext{\textit{$^{\ast}$~Department of Chemistry, Purdue University, USA; E-mail: kais@purdue.edu}}
\footnotetext{\textit{$^{\dag}$~Department of Physics and Astronomy, Purdue University, West Lafayette, USA }}
\footnotetext{\textit{$^{\ddag}$~Fritz-Haber-Institut der Max-Planck-Gesellschaft, Faradayweg 4-6, D-14195 Berlin, Germany }}
\footnotetext{\textit{$^{a}$~Purdue Quantum Science and Engineering Institute, Purdue University, West Lafayette, USA }}




\section{Introduction}
Entanglement is a characteristic feature of quantum mechanics\cite{entanglementorigin,entanglesummary}, responsible for quantum non-locality and the violation of Bell's inequality\cite{bellinequality}. 
Entanglement has been an essential resource for the development of quantum information techniques and technologies\cite{shoralgorithm,entanglement1,entanglement2,entanglement3,entangle10,entangle11,entangle12,entangle13,entangle14,entangle15}. Harnessing entanglement for quantum information processing relies on the ability to manipulate quantum systems, whether in the gas or solid phase. 

In our previous work, we investigated entanglement as well as prospects for quantum computing in arrays of optically trapped polar and/or paramagnetic molecules whose Stark or Zeeman levels served as qubits \cite{entangle16,entangle15}. Herein, we consider a Bose-Einstein condensate (BEC)\cite{bec} of $^{87}$Rb atoms confined in an optical trap and investigate the entanglement between its spin and momentum degrees of freedom. The hyperfine Zeeman levels of the atoms together with their quantized momenta may serve as qubits and even higher dimensional qudits i.e. quantum bits with d dimensions.  

We note that the achievement of Bose-Einstein condensation in gaseous systems followed by the demonstration of spin-orbit coupled BECs \cite{lin2011spin} opened up new avenues for quantum control. In the context of reaction dynamics, spin-orbit coupling has been employed in a photo-chemical reaction of BECs \cite{coherentpa} for preparing the reactants in a coherent quantum superposition of multiple Raman-coupled spin components. Thus coherent control of reactants via spin-orbit coupling could serve as a means to control chemical reactions.The spin-orbit coupling in cold atom systems was also studied in the context of synthetic gauge fields\cite{socreview,socreview2}.

This paper is structured as follows. Section 2 introduces the method of combining Raman lasers and radio-frequency fields to generate entanglement between spin and momentum degrees of freedom in a BEC.  Subsequently, it shows how to quantify the entanglement achieved by von Neumann entropy and concurrence. Section 3 discusses the entanglement  of the spin-momentum coupled BEC under different coupling schemes as a function of the coupling strengths. Section 4 summarizes the presented results and outlines future research directions.

\section{Theory}

\subsection{The Hamiltonian and its eigenstates}

\begin{figure}
    \centering
    \includegraphics[height=4cm]{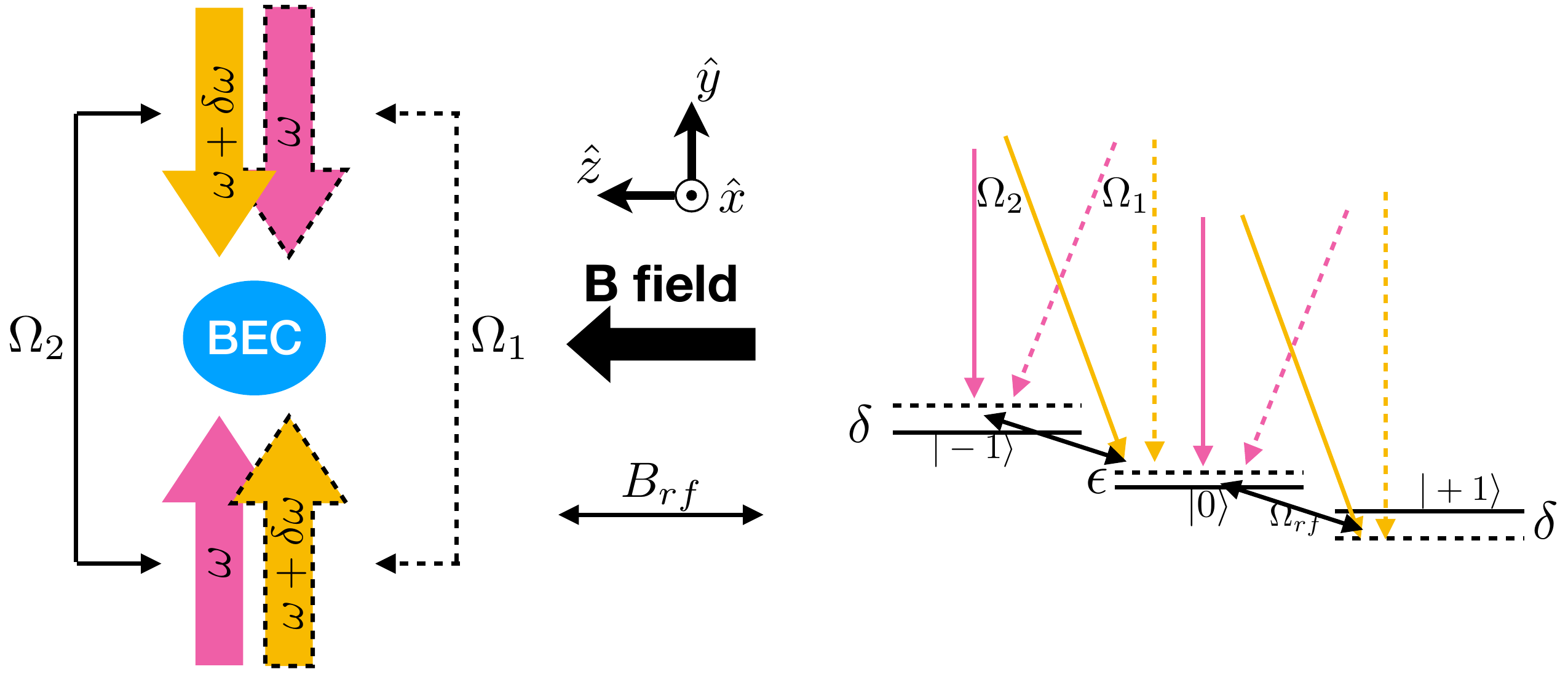}
    \caption{Three mechanisms including two types of Raman laser coupling ($\Omega_{1}$ and $\Omega_{2}$) and a rf field coupling ($\Omega_{rf}$) are applied to drive transitions between different hyperfine spin states.The Zeeman level diagram of a $^{87}$Rb atom in a weak magnetic field. The hyperfine state of $^{87}$Rb in the $F=1$ manifold is split into three Zeeman levels. }
    \label{fig1}
\end{figure}

In order to create spin-momentum coupling in a system of ultracold atoms, 
two counter-propagating Raman lasers were applied to  drive transitions between atomic Zeeman levels \cite{socreview,socreview2}. Experimental features of such a coupling mechanism can be visualized from Figure \ref{fig1} by switching off a Raman $\Omega_{2}$ and the rf coupling $\Omega_{rf}$. As a result, the Rb atom makes a transition from a hyperfine spin state $m_f$ to an $m_f-1$ hyperfine Zeeman state by absorbing and emitting a photon. This process induces, in addition, a change in the momentum of the atom by $2k_r$, where $k_r$ is the photon recoil momentum. To a great extent the coupling scheme is similar to that in Refs. \cite{lin2009bose,lin2011spin}.
The Hamiltonian that describes such a spin-momentum coupling can be written in the coupled basis $\ket{m_{f},K}=\{\ket{-1,q+2K_{r}},\ket{0,q},\ket{+1,q-2k_{r}}\}$ as:
\begin{equation}
    H_0=\begin{pmatrix}
    \frac{\hbar^2}{2m}(q+2k_r)^2-\delta && \frac{\Omega_{1}}{2} && 0 \\
    \frac{\Omega_{1}}{2} && \frac{\hbar^2}{2m}q^2-\epsilon(B) && \frac{\Omega_{1}}{2} \\
    0 && \frac{\Omega_{1}}{2} && \frac{\hbar^2}{2m}(q-2k_r)^2+\delta
    \end{pmatrix}
    \label{eq1}
\end{equation}

 Where $m$ is the mass of $^{87}Rb$, $q$ the quasi-momentum (usually at the minimum of the BEC's lowest energy band), $\Omega_{1}$ the strength of the Raman coupling (which determines the Rabi frequency for the Raman transition between two hyperfine states), $\delta$ the detuning of the Raman laser, $\epsilon(B)$ the quadratic Zeeman shift, and $B$ the strength of the external magnetic field.  In deriving Hamiltonian \eqref{eq1}, we made use of the rotating-wave approximation.
We assume that the BEC is initially created in the $m_f=0$ state. Upon adiabatic turn-on of the Raman coupling, the population is transferred to $m_f=1$ and $m_f=-1$. As a result, the BEC ends up in the ground state of Hamiltonian \eqref{eq1}, which is given by 
\begin{equation}
    |\psi_0\rangle=C_{-1}|q+2k_r,-1\rangle+C_0|q,0\rangle+C_1|q-2k_r,1\rangle
    \label{eq2}
\end{equation}

Hamiltonian \eqref{eq1} only includes one type of spin-momentum coupling, namely $\Omega_{R1}$ that is due to the Raman laser. As it obeys the selection rule $(\Delta m_f=\pm 1, \Delta p=\mp 2k_r)$, many spin-momentum combinations are left uncoupled. In order to couple all spin-momentum states of the BEC, we introduce, in addition, another type of Raman coupling ($\Omega_{R2}$) as well as a coupling by a radio-frequency (rf) field. Fig. \ref{fig1} shows the corresponding Zeeman level diagram for the $F=1$ manifold of $^{87}$Rb together with the three coupling mechanisms that drive transitions between different hyperfine Zeeman states (spin states for short). The selection rule for type 2 Raman coupling is  $(\Delta m_f=\pm 1, \Delta p=\pm 2k_r)$. The rf field only couples different hypefine states, without affecting the momentum of the atoms. Thus, the selection rule for the rf transitions is  $(\Delta m_f=\pm 1, \Delta p=0)$. For convenience, we assume that all couplings have the same detuning, $\delta$. As a result, the Hamiltonian in the spin-momentum basis $|p, m_f\rangle (p=q+2nk_r(n=-\infty\dots\infty), m_f=-1,0,1)$ can be written as
\begin{equation}
\begin{split}
H=\sum_{p,m_f}\left(\frac{p^2}{2m}+m_f\delta+\epsilon(B)\delta_{m_f,0}\right)|p, m_f\rangle\langle p, m_f|+\\\sum_{p,m_f,m_{f'}}\frac{\Omega_{rf}}{2}\delta_{m_f,m_{f'}+1}|p, m_f\rangle\langle p, m_{f'}|+\\\sum_{p,p',m_f,m_{f'}}\frac{\Omega_{R1}}{2}\delta_{p,p'+2k_r}\delta_{m_f,m_{f'}-1}|p, m_f\rangle\langle p', m_{f'}|+\\
\sum_{p,p',m_f,m_{f'}}\frac{\Omega_{R2}}{2}\delta_{p,p'+2k_r}\delta_{m_f,m_{f'}+1}|p, 
m_f\rangle\langle p', m_{f'}|+\text{h.c.}
\end{split}
\label{eq3}
\end{equation}
where the second term represents the coupling induced by the rf field, the third term the coupling by the Type 1 Raman laser and the fourth term the coupling by the Type 2 Raman laser. Upon adiabatic turn-on of the couplings, the ground state of the spin-momentum coupled BEC becomes
\begin{equation}
|\psi\rangle=\sum_{n=-\infty}^{+\infty}\sum_{m_f=-1}^{+1}C_{n,m_f}|q_0+2nk_r,m_f\rangle
\label{eq4}
\end{equation}

\subsection{The entanglement and its characterization}

Entanglement describes the degree of correlation between two or more subsystems of a quantum system. 
It is invariant under local unitary operations on either of the subsystems and under any classical communications between/among them (LOCC)\cite{quantuminfoscience}. In what follows, we quantify entanglement either by (a) von Neumann entropy\cite{entropy1,entropy2,entropy4} or by (b) concurrence\cite{concurrence1,concurrence3,concurrence4}.

\medskip
\noindent \emph{Characterization of entanglement by von Neumann entropy}
\medskip

By making use of the Schmidt decomposition, the ground-state wavefunction, Eq. \eqref{eq4}, can be recast in the form
\begin{equation}
    |\psi\rangle=\sum_{i=1}^d\alpha_i|u_i\rangle_A|v_i\rangle_B
    \label{eq5}
\end{equation}
where $\alpha_i$ is the Schmidt coefficient, $|u_i\rangle$, and $|v_i\rangle$ are the orthonormal vectors of the two subsystems designated as $A$ and $B$.Here the subsystem $A$ and $B$ corresponds to hyperfine spin and momentum respectively of the SOC BEC. The number of terms in this decomposition is $d\le \text{min}\{dim\mathcal{H}_A, dim\mathcal{H}_B\}$, with $dim\mathcal{H}_A$ and $dim\mathcal{H}_B$ the dimensions of the Hilbert space of the subsystems $A$ and $B$, respectively. The number of terms $d$ is also called the Schmidt number. 
For a bipartite pure system, such as $A+B$, the von Neumann entropy is defined as
\begin{equation}
    S=-\sum_{i=1}^d |\alpha_i|^2 \text{Log}_{3}|\alpha_i|^2,
    \label{eq6}
\end{equation}

The model we have considered here is a pair of qutrits where one qutrit corresponds to three momentum states $\{\ket{q+2K_{r}},\ket{q},\ket{q-2k_{r}}\}$ and the second qutrit corresponds to three hyperfine spin states $\{\ket{-1},\ket{0},\ket{+1}\}$. Since we are calculating the Von Neumann entropy of a pair of qutrit we have to take the base of the log as 3.
Information theory tells us that the von Neumann entropy measures how random the subsystem $A$ is when ignoring all information on the subsystem $B$. Thus, maximal entropy is attained when the probability of all possible states in one of the two subsystems is the same. 

As for our $^{87}$Rb BEC with spin-momentum coupling, maximum entanglement and hence maximum von Neumann entropy is attained when the population is equally distributed over the three hyperfine spin states, which results in an entropy of $S_{\text{max}}={\log}_{3}(3)$. 

\medskip
\noindent \emph{Characterization of entanglement by concurrence}
\medskip

The entanglement of formation, which characterizes the amount of entanglement needed in
order to prepare a state described by a density matrix, $\rho$, can be also quantified by concurrence, a quantity that ranges between 0 (no entanglement) and 1 (perfect entanglement). 
 
The evaluation of concurrence is well established for a bipartite systems of two or many qubits \cite{concurrence1,concurrence3,concurrence4}. Typical steps involved in the evaluation are: (i) calculating the density matrix $\rho = \ket{\psi_{0}} \bra{\psi_{0}}$; (ii) constructing the flipped density matrix $\widetilde{\rho} = (\sigma_{y} \bigotimes \sigma_{y}) \rho^{*} (\sigma_{y} \bigotimes \sigma_{y}) $, with $\sigma_{y}$ a Pauli matrix; (iii) calculating eigenvalues $\lambda_{1}, \lambda_{2},\lambda_{3} ...$ of $\rho \widetilde{\rho}$; (vi) retrieving the concurrence $C(\rho)$ from  $C(\rho) = \text{max}[0,\sqrt{\lambda_{1}}-\sqrt{\lambda_{2}}-\sqrt{\lambda_{3}}-...]$.  As the system considered is a pair of qutrits, we need for the evaluation of concurrence analogs of $\sigma_{y}$ in three dimensions. This can be accomplished by making use of the  the generalized approach of Herreno-Fierro and Luthra\cite{herreno2005generalized}. 

\section{Results and discussions}

In this section, we present and discuss the entanglement between the spin and momentum degrees of freedom in an $^{87}$Rb BEC that is spin-momentum coupled according to three different schemes. In our calculations, we use the experimental parameters of Ref. \cite{coherentpa}.

\subsection{A BEC spin-momentum coupled by a single Raman field} 

A comparison of Eqs. \eqref{eq2} and \eqref{eq5} reveals that the ground state wave function is already in the Schmidt-decomposition form, with $C_{-1}$, $C_{0}$ and $C_{1}$ the Schmidt coefficients. In order to maximise entanglement, we have to make the distribution over the three hyperfine spin states as even as possible. As we know, the distribution can be ``tuned'' by varying the Raman coupling strength and the detuning.

\begin{figure}[h]
    \centering
    \includegraphics[height=4cm]{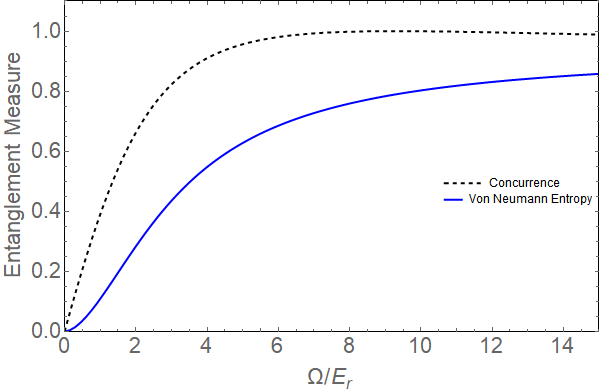}
    \caption{Entanglement between the spin and momentum degrees of freedom for the coupling scheme and parameters of Ref.\cite{coherentpa}. Entanglement measures are shown as function of Raman coupling strength/recoil energy at zero detuning. }
    \label{entrop_omega}
\end{figure}

\begin{figure}[h]
    \centering
    \includegraphics[height=4cm]{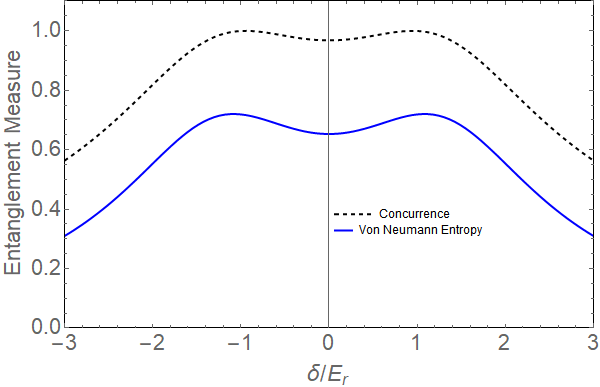}
    \caption{Entanglement properties in the experiment of Ref.\cite{coherentpa}. Entanglement between the spin and momentum degrees of freedom using the coupling scheme and parameters in Ref.\cite{coherentpa}. Entanglement measures as a function of Raman laser detuning/recoil energy at coupling strength $\Omega=5.4 E_r$. The von Neumann entropy, $S$, is divided by its maximum attainable value, $S_{\text{max}}$, see text. }
    \label{entrop_del}
\end{figure}
Figs. \ref{entrop_omega} and \ref{entrop_del} show, respectively, how the entanglement measures -- von Neumann entropy and concurrence -- depends on the Raman coupling strength and the detuning of the Raman lasers. The external magnetic field is $B=5G$, resulting in a quadratic energy shift of $\epsilon=0.65~E_r$, with $E_r=\hbar^2k_r^2/2m$ the recoil energy. 

At zero Raman coupling in Fig \ref{entrop_omega}, the BEC is initially in the $m_f=0$ state with no correlation between spin and momentum. With increasing Raman coupling strength, the other two spin states, $m_f=-1$ and $m_f=1$,  begin to get populated, as a result of which the Von Neumann entropy steeply increases and then saturates -- at about $0.87$ for the parameters used. The concurrence follows suite.  In fact, even when the Raman coupling is quite large and the quadratic Zeeman shift small enough, at most half of the population in the $m_f=0$ state can be transferred to the other two hyperfine spin states (for the adiabatic case), that is, $|C_{-1}|=|C_{1}|=1/\sqrt{2}|C_{0}|$. The maximum entanglement that can be attained for this scheme is $S=(3/2)\text{Log}_{3}2$. Fig. \ref{entrop_del} shows the relative von Neumann entropy and concurrence as functions of the Raman laser detuning at a coupling strength $\Omega_{1}=5.4~E_r$. It is noteworthy that the Von Neumann entropy reaches a maximum of 0.72 (at $\delta\approx \pm 1.06~E_r$), where the populations of the $m_f=1$ and $m_f=-1$ states are unequal.

\begin{figure}[h]
    \centering
    \includegraphics[height=4cm]{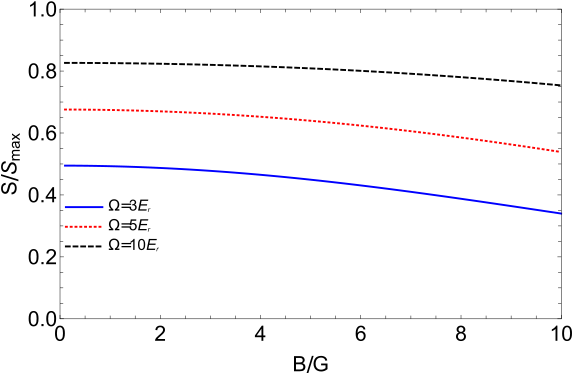}
    \caption{The relative entanglement entropy, $S/S_{\text{max}}$, as a function of the external magnetic field at different Raman coupling strength $\Omega=3,5,10~E_r$ using the scheme of Ref. \cite{coherentpa}. All Raman laser detunings are zero.}
    \label{fig3}
\end{figure}

Since the external magnetic field affects the hyperfine levels and induces a quadratic Zeeman shift, it is worthwhile to investigate the dependence of the entanglement on the magnetic field. Fig. \ref{fig3} shows the dependence of the relative von Neumann entropy on the magnetic field strength for several values of the Raman coupling strength. In this calculation, the Raman frequency is tuned to resonance with transitions between different spin states, i.e., the detuning $\delta=0$. The von Neumann entropy is seen to decrease as the quadratic Zeeman shift becomes large, but this effect becomes less significant at large coupling strengths. The above coupling scheme was studied before, see Refs.\cite{youprb,zhao2019,liucpb}. For instance, Ref. \cite{zhao2019} dealt with the entanglement dynamics of a spin-1 BEC. Ref. \cite{liucpb} tackled the enhancement of spin-momentum entanglement in classical chaos.

The above coupling scheme only allows us to access three states in the spin-momentum Hilbert space. In order to make better use of the spin-momentum coupled BEC in quantum information processing, we introduce two additional schemes that make it possible to access many more or even all possible states in the spin-momentum Hilbert space.

\subsection{A BEC spin-momentum coupled by one type of Raman field and an rf field} 

\begin{figure}[h]
    \centering
    \includegraphics[width=0.4\textwidth]{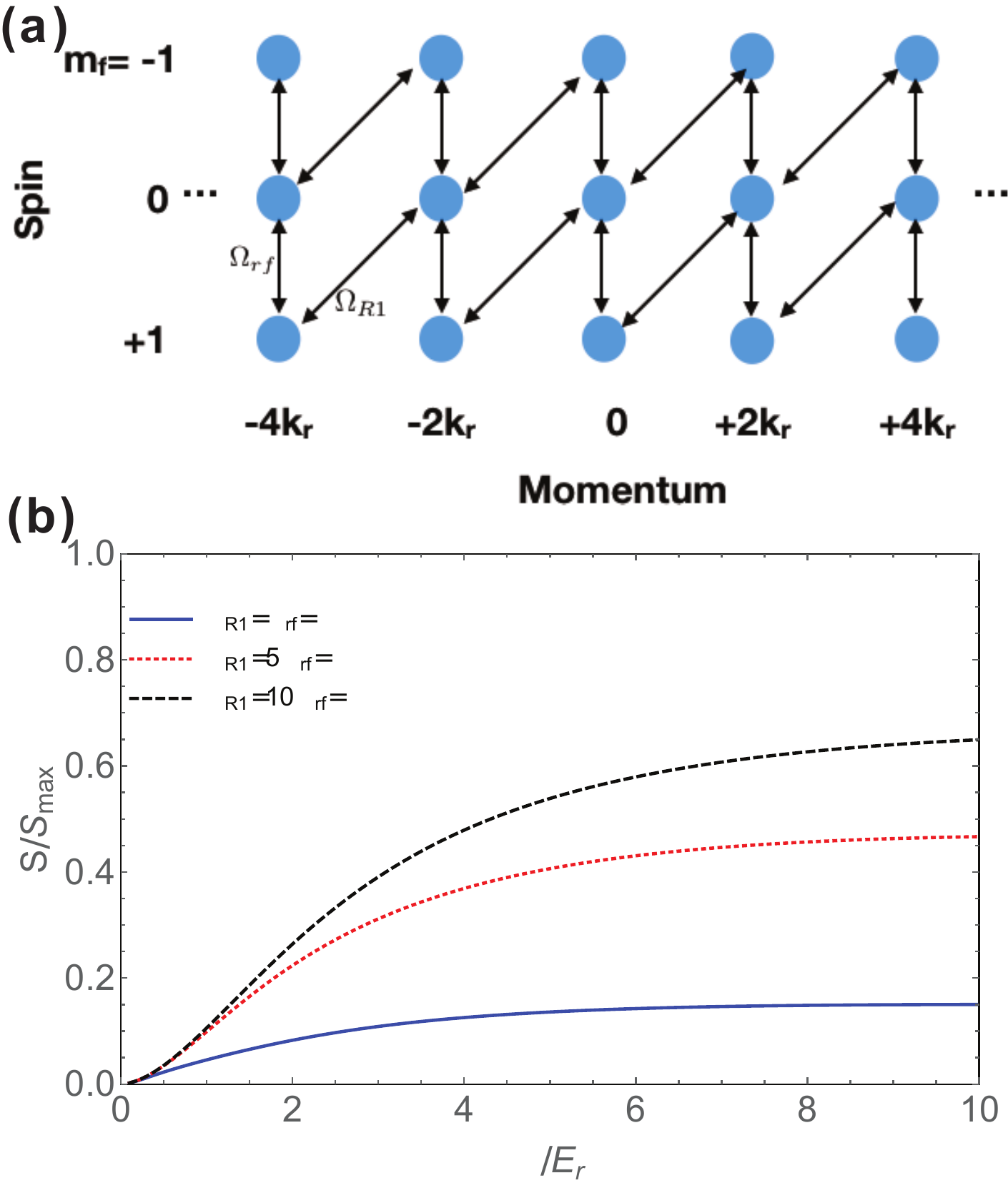}
    \caption{(a) Creating entanglement between spin and momentum degrees of freedom of a BEC using combined Raman and rf fields. The Raman laser always couples states with different spin and momentum, with the selection rule $(\Delta m_f=\pm 1, \Delta p=\mp 1)$. The rf field flips spins of the atoms without affecting the momentum $(\Delta m_f=\pm 1)$. (b) The relative entanglement entropy as a function of the coupling strength. The variation of the relative entanglement entropy is shown for various ratios of the Raman-to-rf coupling strengths.}
    \label{fig4}
\end{figure}

The scheme that makes use of coupling by one type of Raman field in combination with an rf field is shown in Fig. \ref{fig4}(a). Moreover, the experimental features of such a coupling mechanism can be visualized from Figure \ref{fig1} by switching off $\Omega_{2}$. This scheme was employed in an experiment to create a lattice potential for cold atoms\cite{latticepotential}. Any spin-momentum states $|p, m_f\rangle$ with $p=q+2nk_r$, $n=-\infty\dots\infty$, and $m_f=-1,0,1$ are accessible within this scheme and mutually correlated. Assuming that the BEC is created adiabatically in the spin-momentum superposition state, Eq. \eqref{eq4}, we evaluated the entanglement between spin and momentum for different Raman and rf coupling strengths at zero detuning, see Fig. \ref{fig4}(b). 

The entanglement is seen to increase with the coupling strength but then saturates for $\Omega \geq 10~E_r$. The entanglement increases significantly when the ratio between the Raman and rf coupling $\Omega_{R1}/\Omega_{rf}$ becomes large. This is because the rf field only induces transitions between different spin states without affecting the momentum of the atom. As more spin states are populated, the system's wave function becomes increasingly separable.

Although all momentum states are coupled by the combined Raman and rf transitions, an adiabatic population transfer from small to large momentum states entails higher-order effects as the atom must undergo multiple transitions.
The experiment reported in Ref. \cite{latticepotential} shows that for $\Omega_{R1}\approx 10~E_r$ there is a detectable population in $|p|\le 4k_r$ states. 
This finding is reproduced by our calculation of the von Neumann entropy, as shown in Fig. \ref{fig5}. The entropy is found to converge for $|p|$ up to $|p|\le 4k_r$ for $\Omega_{R1}=10~E_r$. Higher momentum states are not populated.

\subsection{A BEC spin-momentum coupled by two types of Raman fields} 

\begin{figure}[h]
    \centering
    \includegraphics[width=0.4\textwidth]{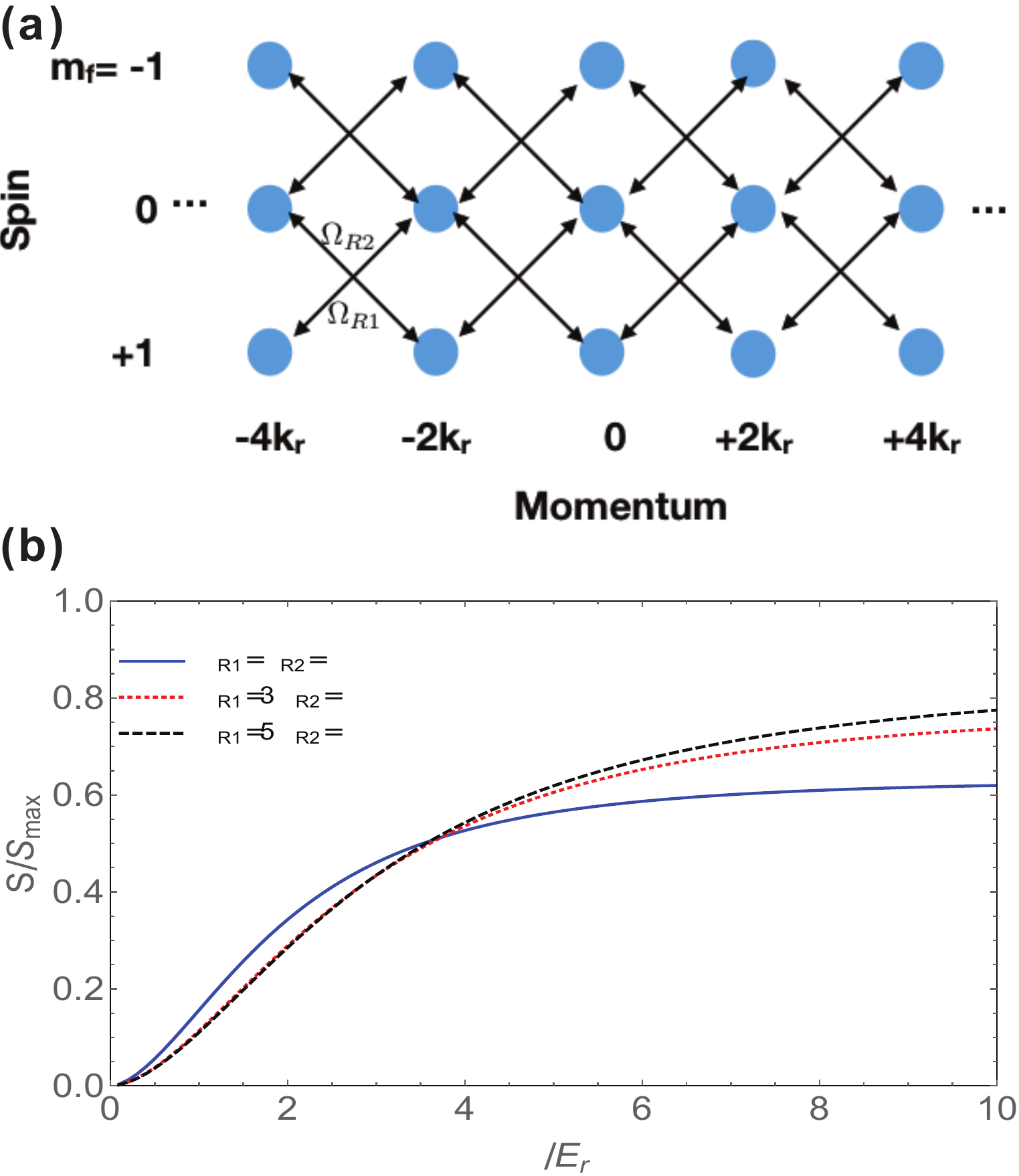}
    \caption{(a) Creating entanglement between spin and momentum degrees of freedom of a BEC using two types of Raman laser couplings. Type 1 Raman coupling has the selection rule $(\Delta m_f=\pm 1, \Delta p=\mp 1)$. Type 2 Raman coupling has the selection rule $(\Delta m_f=\pm 1, \Delta p=\pm 1)$. (b) The relative entanglement entropy as a function of the coupling strength at zero detuning for both Raman lasers. The dependence of the von Neumann entropy is shown for different ratios of the two Raman coupling strengths.}
    \label{fig6}
\end{figure}

\begin{figure}[!]
    \centering
    \includegraphics[height=4cm]{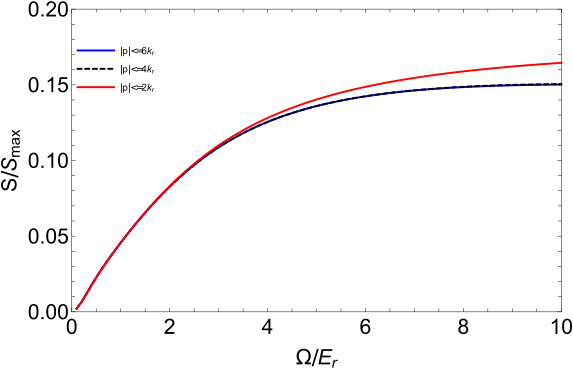}
    \caption{Relative entanglement entropy as a function of the coupling strength for the combined Raman and rf coupling mechanisms. The calculations took into account different numbers of momentum states. For the coupling strength of up to $\Omega=10~E_r$, convergence is reached at $|p|\le 4k_r$. The coupling strength is $\Omega_{R1}=\Omega_{rf}=\Omega$. All detunings were set to zero.}
    \label{fig5}
\end{figure}

Multiple states in the spin-momentum Hilbert space can also be accessed by combining two different Raman couplings, see  Fig. \ref{fig6}(a). The corresponding two types of Raman couplings have different selection rules, namely $(\Delta m_f=\pm 1, \Delta p=\mp 2k_r)$ and $(\Delta m_f=\pm 1, \Delta p=\pm 2k_r)$. Using these two couplings in conjunction results in transferring population to higher momentum states. However, only half of the states shown in the figure are accessible if the formation of the BEC is initiated at zero momentum and the two Raman couplings are applied adiabatically. States with $m_{f1}=m_{f2}$ and $|p_1-p_2|=2nk_r$ with $n=1,3,5,\dots$ are not coupled in this scheme. Fig. \ref{fig6}(b) shows the relative entanglement entropy as a function of the coupling strength of the two Raman coupling mechanisms at zero detuning. It follows that large entanglement between spin and momentum can be obtained at $\Omega_{R1}=10~E_r$. A bias of the coupling strength -- such that $\Omega_{R1}/\Omega_{R2}>1$ -- increases the entanglement, but not as significantly as the combined Raman and rf coupling scheme.

\section{Conclusion and outlook}
We examined the entanglement between spin and momentum degrees of freedom in an optically trapped $^{87}$ Rb BEC that arises due to coupling by combined Raman and rf fields. All states of the spin and momentum Hilbert space can be accessed by combining two types of Raman coupling with coupling by an rf field; a multitude of spin-momentum states can also be accessed without applying the rf field. As these spin-momentum states could serve as a computing unit (a pair of qutrit), we evaluated the entanglement between spin and momentum degrees of freedom. Our calculations indicate that a large entanglement -- as quantified in terms of von Neumann entropy and concurrence -- can be achieved at feasible values of the coupling strengths \cite{macroscopic}.

Although the individual atoms in a BEC are indistinguishable, different BECs are distinguishable. Their macroscopic quantum nature was used in quantum computing with collective logical encoding which increases the performance by reducing the time required for two-qubit operations by a factor of N, with N the number of atoms in each of the BECs \cite{macroscopic}. Also, a fully single-qubit control on the Bloch-sphere has been achieved for two-component BECs realized on atom chips \cite{coherent_atomic_chip}.

Next, we plan to investigate the creation and control of entanglement between spatially separated parts of a BEC\cite{becspatial,becspatial2}. We will also explore the analogy with the system of trapped ions that has been implemented for quantum computing\cite{iontrapqc}. 
Finally, we will examine the role of entanglement in controlling and predicting the quantum interference patterns that have been observed in scattering experiments such as those on the photo-association reaction of spin-momentum coupled BEC \cite{coherentreaction1,coherentreaction2,coherentreaction3}.

\section*{Acknowledgements} This work is supported by the U.S. Department of Energy, Office of Basic Energy Sciences, under award number DE-SC0019215. We thank Esat Kondakci and Chuan-Hsun Li for useful discussions.This article is a part of themed collection:\href{https://pubs.rsc.org/en/content/articlelanding/2020/cp/d0cp03945d/unauth#!divAbstract}{Quantum Computing and Quantum Information Storage}



\balance


\bibliography{template} 
\bibliographystyle{rsc} 

\end{document}